\documentclass[a4paper,11pt]{article}
\pdfoutput=1 

\usepackage{jheppub} 

\usepackage[T1]{fontenc} 

\title{\boldmath Gravitational correlation, black hole entropy and information conservation}

\author[a]{Dongshan He}
\author[b]{Qing-yu Cai \note{Corresponding author.}}

\affiliation[a]{College of Physics \& Electronic Engineering, Xianyang Normal University. Xianyang 712000, China}

\affiliation[b]{State Key Laboratory of Magnetic Resonances and Atomic and Molecular Physics,
Wuhan Institute of Physics and Mathematics, Chinese Academy of Sciences, Wuhan
430071, China}

\emailAdd{hfrnsm@163.com}
\emailAdd{qycai@wipm.ac.cn}

\abstract{When two objects have gravitational interaction between them, they are no longer independent of each other. In fact, there exists gravitational correlation between these two objects. Inspired by E. Verlinde's paper\cite{ev11}, we first calculate the entropy change of a system when gravity does positive work on this system. Based on the concept of gravitational correlation entropy, we prove that the entropy of a Schwarzschild black hole originates from the
gravitational correlations between the interior matters of the black hole. By analyzing the gravitational correlation entropies in the process of Hawking radiation in a general context, we prove that the reduced entropy of a black hole is exactly carried away by the radiation and the gravitational correlations between these radiating particles, and the entropy or information is conserved at all times during Hawking radiation. Finally, we attempt to give a unified description of the non-extensive black-hole entropy and the extensive entropy of ordinary matter.
}


\begin{document}
\maketitle

\section{Introduction}

In 1971, Hawking showed that the total area of the event horizon of a
classical black hole can never decrease\cite{swh71}. This result, known as the classical area law, is remarkably similar to the second law of thermodynamics. Bekenstein then
proposed that a black hole should have an entropy, and it should be
proportional to the horizon area of the black hole\cite{jdb73}.
Soon after that, Hawking showed that a black hole will have thermal Hawking radiation and
therefore possess a temperature\cite{swh74,swh75}. However, this
discovery brings up a series of problems. Irrespective of what initial
state a black hole starts with before collapsing, it evolves eventually into
a thermal state after being completely evaporated into emitted radiations.
It is well known that information is physical and thus it is conserved
during any physical process\cite{rl91,nc00}. But, Hawking radiation violates
information conservation or equivalently entropy conservation.
This is the so-called paradox of black hole information loss\cite{swh76}.

Many approaches have been suggested for resolving this paradox of black hole information loss(see, for example, the recent reviews \cite{BC11,Har14}). Notably, based on a
non-thermal spectrum for the Schwarzschild black hole, one of us(Qing-yu Cai) with his collaborators, has recently proved that the black hole radiation is an entropy
conservation process by counting the entropy taken out by the emitted
particles\cite{zcyz09}.

Another mystery about a black hole is the origin of its  entropy.
This is also important for solving the paradox of black hole information loss.
Some authors (see, for example, \cite{BKLS86,ms93,fn93}) have
shown that the entropy of a black hole is the entanglement entropy of the entangled fields  across the black hole horizon. However, general consensus on this point have not been reached yet. In 2010, E. Verlinde's in his remarkable paper\cite{ev11} explained gravity as an entropic force. This new point of view has aroused extensive discussions. It relates  gravity and
the entropy change of a system through the thermodynamic method,
with which it is easy to calculate the entropy change of a
system when the system interiors have gravitational interactions between them.

In this paper we will, with the help of E. Verlinde's entropic gravity,  introduce the concept of gravitational correlation entropy to solve the above problems. The correlation entropy can be easily understood  already in ordinary cases.
In ordinary thermodynamics entropy is an extensive quantity. If there are no
interaction between system $A$ and $B$, then the thermal entropy of the total system $A+B$
is $S(A,B)=S(A)+S(B)$. If we include the interactions between $A$ and $B$, then
$A$ and $B$ is no longer independent of each other, and $S(A,B)=S(A)+S(B)+S_c(A:B)$, where $S_c(A:B)$ is the correlation entropy. In quantum information theory, $-S_c(A:B)$ is the mutual information between $A$ and $B$\cite{nc00}, which is a legitimate measure for the total amount of correlations between two parts $A$ and $B$ of a bipartite system $A+B$.

In the following, we first calculate the gravitational correlation entropy $S_c(M:m)$
between two objects (with respective masses) $M$ and $m$ under some mild assumptions. Then by analysing the formation process of a black hole, we effectively envision that an object $m$ will first self-collapse to form a small black hole before it is combined with $M$ to form a black hole $M+m$, so as to calculate the entropy change on both holographic screens of $m$ and $M$. In this way all the increased entropy of the black hole is the gravitational correlation
entropy. Through further analysis, we prove that the black hole entropy originates
from the gravitational correlations between the interior matters that formed the black hole.

Because  there are strong gravitational correlations between the interior matters of a black hole, the emissions of Hawking radiation must be statistically dependent, and correlations
must exist between them. We prove that the entropy is conserved in Hawking radiation
and after a black hole is exhausted by Hawking radiation, all entropy of the
black hole is carried away by radiated particles and the correlations between these particles. We will work in a general context which is independent of the tunneling formalism of Hawking radiation, and hence the method of \cite{zcyz09} is further assured.

Furthermore, since the main feature of the Bekenstein-Hawking entropy is its proportionality to the area of the black hole horizon $S_{BH}\propto A\propto M^{2}$, it is rather
different from the usual entropy, e.g. the entropy of a thermal gas in a box,
which is proportional to the volume $S_{o}\propto V\propto M$.  We will calculate the gravitational correlation entropy of ordinary matters so as to give an attempt on the unification of the non-extensive (black-hole) entropy and the extensive ordinary one. We show that the reason of
the difference between the black hole entropy and the entropy of ordinary matter is  the fact that the correlation entropy in ordinary matter caused by gravitational interactions between ordinary particles is too small to be important.

This paper is organized as follows. In next section, we compare the black hole entropy with the ordinary thermal entropy to introduce the concept of correlation entropy. In section III, we use entropic gravity to show that the gravitational correlation entropy between the interior matter of a black hole is the origin of  black hole entropy. In section IV, by calculating the gravitational correlation entropy between Hawking radiations, we show that information is conserved at all times. In section V, we use correlation entropy to unify the entropies of black holes and ordinary matter. Sec. VI concludes with some remarks.

\section{Correlation entropy}

In statistical mechanics, if we divide an equilibrium system into a number of
macroscopic parts, the total number of microscopic states is a product of the
number of microscopic states $\Omega_{i}$ of each part, i.e. $\Omega=\prod_{i}%
\Omega_{i}$. The Boltzmann entropy $S=k\ln{\Omega}$ is then extensive
$S=\sum_{i}S_{i}$. For example, the entropy of the monatomic
ideal gas is\cite{Lee}
\[
S=\frac{3}{2}Nk\ln T+Nk\ln\frac{V}{N}+\frac{3}{2}Nk\left(  \frac{5}{3}%
+\ln\frac{2\pi mk}{h^{2}}\right)  ,
\]
where $N$ is the number of the atoms. We can see that $S\propto N$, when
the temperature $T$ and the the particle number density $N/V$ are kept constant.

In a realistic system, however, there are all kinds of interactions including the gravitational one, and the dynamics of one particle  is no longer independent of those of other particles.
In this case, if we divide successively an equilibrium system into
several subsystems, then to a certain length scale $l$, the physics
of a subsystem might be no longer the same as the original one. The extensive
property of the thermal entropy $S(A,B)=S(A)+S(B)$ is then violated\cite{gsl08}.

The entropy of a Schwarzschild black hole (with mass) $M$ is
$S(M)=4\pi M^{2}$. When we divide a black hole with mass $M+m$
into two parts (with respective masses) $M$ and $m$, we have%
\begin{align}
S_{BH}(M+m) &  =4\pi(M+m)^{2}\nonumber\\
&  =4\pi M^{2}+4\pi m^{2}+8\pi Mm\label{sMm2}\\
&  \neq S_{BH}(M)+S_{BH}(m).\nonumber
\end{align}

The third term in the second line $8\pi Mm$ measures some kind of correlation.
We called this term $S_{c}(M,m)=8\pi Mm$ the correlation entropy of a Schwarzschild black hole.

In contrast to the usage of mutual information in \cite{zcyz09,gsl08}, we have defined the {\it gravitational} correlation entropy as the negative of the original mutual information. This is can be motivated from the following observations. For the case of black hole coalescence, $S_{BH}(M+m)>S_{BH}(M)+S_{BH}(m)$ by the classical area law\cite{swh71} and hence the mutual information is negative. While in the case of Hawking radiation, the mutual information is positive as is shown in \cite{zcyz09}, which is correct for the disintegration of a black hole into Hawking radiation and a remnant if additional resources are required to store the released information taken out by Hawking radiation.
From the viewpoint of the noiseless channel coding theorem\cite{nc00}, these two processes are complimentary:
\begin{itemize}
\item For black hole coalescence, the total entropy increases and the original information are combined into a single black hole, which means inside the black hole the increases in gravitational correlations are used as resources to store such new information. Hence the mutual information should be positive and we need a negative sign in front of the usual definition.
\item For Hawking radiation, the black hole entropy is reduced, which means the gravitational correlations inside the black hole are less required to store the remaining information. Hence the mutual information could be negative. That the mutual information calculated in \cite{zcyz09} is positive is because  in that case the resources are required {\it outside} the black hole.
\end{itemize}
Therefore, it is adequate to choose the negative of the usual mutual information as the measure of gravitational correlation inside a black hole.

Now a question is how to study the correlation {\it inside} the black hole horizon. This is almost impossible since our understanding of the interior of a black hole is very little. In the following, we use an effective way to calculate these correlations by studying the entropy change during the formation of  a black hole.

\section{Gravitational Correlation entropy}

Inspired by Bekenstein's entropy bound, E. Verlinde postulated that when a test
particle (with mass) $m$ moves towards a holographic screen, the entropy on the holographic
screen will increased by $\Delta S\sim2\pi m\Delta x$. This entropy change leads
to a force which is called entropic force, and gravity is explained as an
entropic force\cite{ev11}. There is still a debate about this point of view.
Here we suggest that the increased entropy is a result of gravity. That is to say, the entropy change on the holographic screen $ \Delta S\sim2\pi m\Delta x $ originates from the work done by gravity (which has already been pointed out in \cite{Gao11} as an objection to the interpretation of entropic gravity).

Let us first recall some basic elements of entropic gravity. Consider two objects (with respective masses) $M$ and $m$, and let $F$ be the interaction
force between them. With $M$ being the reference system, according to the second
law of Newton, we have%
\[F=ma,\]
where $a$ denotes the acceleration of object $m$. It is well known that the
acceleration $a$ and the temperature $T$ are closely related by Unruh
effect. Namely, an observer in an accelerated frame experiences a temperature\cite{wu76}%
\begin{equation}
T=\frac{\hbar|a|}{2\pi kc}. \label{unruh}
\end{equation}
where $k$ is the Boltzmann constant, $c$ the speed of light, and $a$ the acceleration of the frame.
If  the direction of the x-axis of the accelerating frame is from the holographic screen of $M$ to the test
particle $m$, then $a<0$ and there will be a change of sign in comparison to \cite{ev11}. When the test particle $m$ approaches the object $M$, the energy of the system
$M$ did not change, namely $dE_{M}=0$. According to the first law of thermodynamics, we have the entropic force relation on the holographic screen\cite{ev11}
\begin{equation}
F\Delta x=T\Delta S, \label{ther}%
\end{equation}
where $T$ is the temperature on the screen induced by Unruh effect. Then we can get%
\begin{equation}
\Delta S_{scr}=-2\pi k\frac{mc}{\hbar}\Delta x \label{ds/dx}%
\end{equation}
for a plane screen or when the test particle $m$ is very close to the screen.
When the test particle is not very close to screen, from the Newton's law of
gravity
\[
F=G\frac{Mm}{x^{2}},
\]
we can get a generalized form of (\ref{ds/dx}). For a spherical screen with radius
$r_{scr}$, when the test particle $m$ is at the distance $x$ form the center of
ball bounded by the spherical screen, we have the generalized entropic force relation
\begin{equation}
\Delta S_{scr}=-2\pi k\frac{mc}{\hbar}\left(  \frac{r_{scr}}{x}\right)
^{2}\Delta x. \label{ds/dx1}%
\end{equation}
When the test particle is very close to holographic screen $r_{scr}%
/x\rightarrow1$, then Eq.(\ref{ds/dx1}) returns to Eq.(\ref{ds/dx}). The
Eq.(\ref{ds/dx1}) shows that the interaction between the screen and the test
particle will be weak if the test particle becomes far away from screen.

In this section we will use Eq.\eqref{ds/dx1} to define and calculate the gravitational correlation entropy of a black hole. Form now on, we will set $k=\hbar=c=G=1$ for simplicity.

\subsection{The radius of the holographic screen}

In E. Verlinde's article\cite{ev11}, there are no constraints on the radius of the
holographic screen. But he requires that the test particle must be very close to the
holographic screen, so that Eq.(\ref{ds/dx}) is valid. When the test particle moves towards the screen, the radius
of the screen is expected to shrink, that is to say, there is no stationary screen in
E. Verlinde's proposal.

Now using Eq.(\ref{ds/dx1}), we can calculate the increased entropy on
the screen no matter where the test particle is. For example, when the test
particle $m$ is translated from $R_{1}$ to $R_{2}$, ($R_{1},R_{2}>r_{scr}$) and the radius of the holographic screen is independent of the position of the test particle $m$, the entropy on the screen will increase by
\begin{align*}
\Delta S_{scr}  &  =-\int_{R1}^{R_{2}} 2\pi m\left(  \frac{r_{scr}}{x}\right)
^{2}dx=2\pi mr_{scr}^{2}\left(  \frac{1}{R_{2}}-\frac{1}{R_{1}}\right)  .
\end{align*}
When $m$ moves from $R_1$ to $R_2$($R_1$,$R_2$ are constants.), the entropy change $\Delta S_{scr}$ should
be constant. This resembles the unitary requirement for the translation of the radiated particles\cite{Kob10,AN13} where the holographic entropy of $m$ is required to be a constant under such a translation. So the increased entropy on the holographic screen should not depend on the screen radius $r_{scr}$.
Therefore, the radius of the holographic screen of $M$ should be a constant for a given object. This observation will be crucial for our discussions on black hole entropy in next subsection.

In order to determine the position of the holographic screen, let us consider some restrictions of the radius $r_{scr}$. First, we observe that the radius $r_{mat}$ of an ordinary spherical matter $M$ can not be smaller than that of the even horizon radius $r_{Sch}$ of a black hole with the same mass, that is, $r_{mat}\geqslant r_{Sch}$. On the other hand, if  $r_{scr}>r_{Sch}$ then
\[\Delta S_{scr}(\infty\rightarrow r_{scr})=2\pi mr_{scr},\] which will lead to the contradiction that the entropy
change of the ordinary object $M$ is bigger than a black hole with the same mass $M$. Therefore the radius of
the holographic screen $r_{scr}$ should be no greater than the Schwarzschild radius $r_{Sch}$, i.e. $r_{scr}\leqslant r_{Sch}$. Therefore, for objects made from ordinary matters, the $r_{scr}$ is usually inside the object and is bounded by the the Schwarzschild radius, i.e. $r_{scr}\leqslant r_{Sch}\leqslant r_{mat}$. Second, since we are interested in calculating the gravitational correlation entropy, while
the structures inside the inner black hole (of radius $r_{Sch}$) are generally unknown, it is impossible to do calculations if we choose $r_{scr}<r_{Sch}$.
Therefore, we suggest that the holographic screen radius $r_{scr}$ of an object $M$ can be effectively taken as the Schwarzschild radius,\footnote{Of course, the entropy on the holographic screen will still  increase after it falls behind the black-hole horizon, but on the one hand it will not affect the black-hole entropy we observe outside, and on the other hand it must be compensated by the decrease on another screen so that the area law of black hole entropy is not violated.}
\begin{equation}
r_{scr}=r_{Sch}=2M. \label{rscr}%
\end{equation}
If $M$ is a black hole, then the holographic screen coincides with the black-hole horizon, whereas if it is ordinary matter, say a test particle, the holographic screen will be still less than the scale of the ordinary matter even if it takes the value of the Schwarzschild radius. There is no contradiction in both cases. A reason for such a choice is that if the holographic principle holds, then the holographic screen should contain all the information inside the screen. In the case of black holes, it is recently showed in \cite{Zhang15} that the black hole's interior volume is not sufficient to hold all the information(entropy) and all the information should be encoded near the horizon, hence it is adequate to assume that the holographic screen coincides with the black hole horizon if we want to know the information about the correlations between its components.

Under this assumption, Eq.(\ref{ds/dx1}) becomes
\begin{equation}
\Delta S=-8\pi M^{2}m\frac{\Delta x}{x^{2}},
\end{equation}
and we have
\begin{equation}
\Delta S_{scr}\leqslant 2\pi mr_{Sch}=4\pi Mm,\label{888888}
\end{equation}
for any screen. The entropy on one screen has a maximum value only when the radius
of the object equals its Schwarzschild radius (of a black hole), $\Delta S_{scr}=2\pi
mr_{Sch}(M)=4\pi Mm$. Hence, the above assumption on the $r_{scr}$ corresponds to the case of maximum entropy. From the point view of entropic gravity, this means the strongest gravity, i.e. black holes. At this stage, it already can be expected that the strongest gravitational correlation entropy calculated under the above assumption  will be able to explain the entropy of a black hole.

Note that Eq.\eqref{888888} is in contradiction with the bound given in \cite{Sah11} where the minimal radius of the holographic screen is found to be much less than the  Schwarzschild radius. This contradiction arises from the difference in assumptions. In our assumption, we have stopped at the  Schwarzschild radius since below it the gravitational correlations are intractable. However, we have also shown that $r_{scr}\leqslant r_{Sch}$, which does not violate the result of \cite{Sah11} too much.

\subsection{The origin of black hole entropy}

When an object with mass $m$ approaches a black hole with mass $M$ from infinity, the entropy change on the holographic screen of $M$ is
\begin{align}
\Delta S_{M)m}=&-\int_{+\infty}^{2M}8\pi M^2 m\frac{dx}{x^2}\nonumber \\
=&8\pi M^2 m\frac{1}{x}\mid_{+\infty}^{2 M}=4\pi M m.
\end{align}
At the same time, the black hole $M$ approaches to the holographic
screen of object $m$. Then, the  entropy change of this screen is
\begin{align}
\Delta S_{m)M}=&-\int_{+\infty}^{2M}8\pi m^2 M\frac{dx}{x^2}
-\int_{2M}^{2m}8\pi m^2 M\frac{dx}{x^2}, \label{deltaSmM} \nonumber\\
=&8\pi m^2 M\frac{1}{x}\mid_{+\infty}^{2 M}+8\pi m^2 M\frac{1}{x}\mid_{2M}^{2m}, \\
=&4\pi m^2+(4\pi M m-4\pi m^2)=4\pi M m.
\end{align}
According to the first term of Eq.(\ref{deltaSmM}), we can find that
the entropy on the holographic screen of $m$ is first increased by $4\pi m^2$
until the distance between the object $m$ and $M$ is comparable (actually equal) to the radius of  the holographic screen of $M$. After that, it is further translated to hit the screen of $m$, which is the second term of Eq.(\ref{deltaSmM}).

\begin{figure}[ht]
\centering
  \includegraphics[height=4cm]{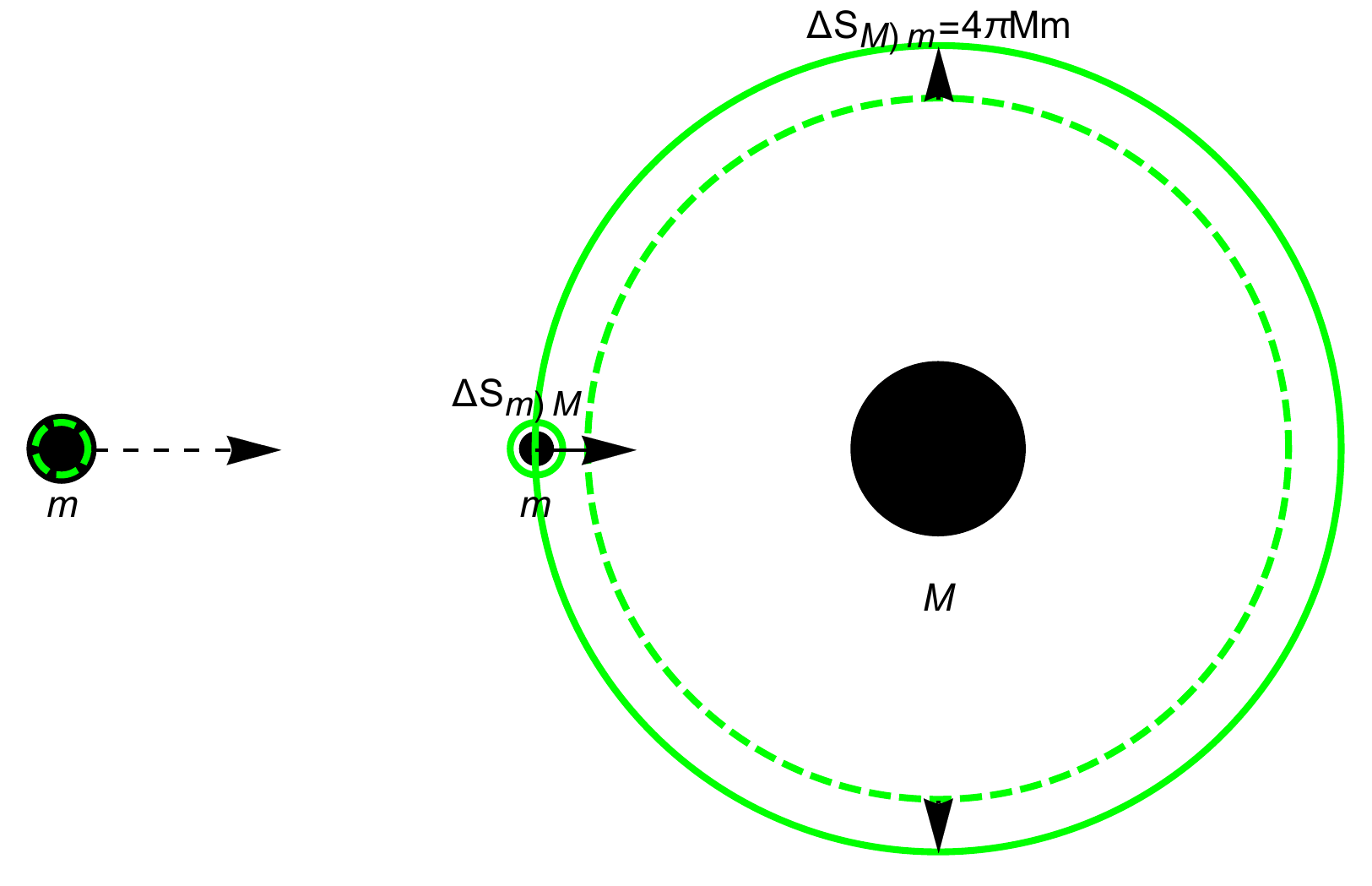}
 \caption{(Color online) A schematic depiction of the process of the formation of a new black hole from a test particle $m$ and a large massive old black hole $M$. The (light) green line represents the holographic screen.}
 \label{mBH}
\end{figure}

Now, since we have taken the radius  of the holographic screen as the Schwarzschild radius \eqref{rscr}, the object $m$  also can  be considered as a (small) black hole of the Schwarzschild radius.
We can envision this virtual process as follows.
The object $m$ first self-collapses to form a small black hole.
Then, the small black hole falls into the old black hole $M$, and the entropy
on the holographic screen of $m$ will increase  during this process but the $r_{scr}$ is kept constant. Finally, two black holes collide to form a new black hole with a new $r_{scr}$. It must be emphasized that the self-collapsing is not a real physical process. The real situation must be that the real radius of $m$ is larger than that of the effective holographic screen. Then the increase in the screen entropy $\Delta S_{m)M}$ will cause the increase in the screen radius continuously and eventually $m$ collides with $M$ forming a new black hole. Such a process is schematically depicted in FIG.1.

In this process, the increased entropy of the system $M+m$ should be the sum of the entropy change on the screen of $m$ and $M$, that is\footnote{If the object $m$ indeed self-collapses to form a small black hole before
it falls into the old black hole $M$. Then due to the fact that the self-gravity of $m$ does
positive work, the self-entropy of $m$ should be increased by $S_c(m)=4\pi m^2.$
The self-entropy of $m$ is a correlation entropy by itself.}
\begin{equation}\label{1515}
  S_{c}(M:m)=\Delta S_{m)M}+\Delta S_{M)m}=8\pi Mm
\end{equation}
which has the same value as in the third term of \eqref{sMm2}.
According to the above analysis, we know that the increased entropy (\ref{1515}) of the system originates from the gravitational interaction between $m$ and $M$. Therefore, we call this quantity  the gravitational correlation entropy between $m$ and $M$. At the formal level, this is an entropy-entropy correlation since the entropy change on the screen of $m$ is correlated with the change on the screen of $M$. In view of entropic gravity, this means exactly the gravitational correlation.

When we consider the formation process of a black hole, all the particles
of the black hole will approach the maximum correlation entropy with respect to other
particles in the interior of the black hole. Assume that a black hole is made up of $N$ particles,
then all the gravitational correlation entropy between them is
\begin{align*}
S_{c}(M)  & =\underset{i=1}{\overset{N}{\sum}}\underset{j=1}{\overset{N}{\sum}}\Delta S_{m_i) m_j}\\
&=\underset{i=1}{\overset{N-1}{\sum}}\underset{j=i+1}{\overset{N}{\sum}}S_c(m_i,m_j)
+\underset{i=1}{\overset{N}{\sum}}S_c(m_i) \\
&=\underset{i=1}{\overset{N-1}{\sum}}\underset{j=i+1}{\overset{N}{\sum}}8\pi m_i m_j
+\underset{i=1}{\overset{N}{\sum}}4\pi m_i^2 \\
& =4\pi M^{2}%
\end{align*}
That is to say, $S_{c}(M)=S_{BH}$, the Bekenstein-Hawking entropy. Thus, we have proved that the entropy of a black hole is indeed the gravitational correlation entropy. In this respect, we can interpret the origin of black hole entropy as the gravitational correlation entropy between particles in the {\it interior} of the black hole that formed the black hole.

\section{Entropy conservation in Hawking radiation}
We have shown that the black hole entropy originates from the gravitational correlations
between the interior matters that formed the black hole. Hawking discovered  that a black
hole excites Hawking radiation \cite{swh74,swh75}. This discovery brings up the paradox of
black hole information loss. In this section, we consider the process of
Hawking radiation from a Schwarzschild black hole and calculate the gravitational correlation entropy between  radiated particles.

First, we consider a Schwarzschild black with mass $M$ and initial entropy  $4\pi M^2$.
When a particle with mass $m_1$ escapes from the black hole, we can calculate
the entropy carried away by the particle. The reduced entropy includes two parts:
the self-entropy of the particle $m_1$ and the correlation entropy of the particle
with the remaining black hole. Then the entropy change of the black hole is
\begin{equation}
\Delta S_1=-S_c({m_1})+S_c(m_1:M-m_1),
\end{equation}
where the first term is $S(m_1)=4\pi m_1^2$, and the second term is the correlation entropy between the emitted particle
and the remaining black hole with mass $M-m_1$. Then we can obtain by Eq.\eqref{1515},
\begin{align}
S_c(m_1:M-m_1)=&-2\int^{\infty}_{2{(M-m_1)}}2\pi m\left[ \frac{2(M-m_1)}{x}\right]^2dx\nonumber\\
=&-8\pi m_1 (M-m_1).\label{17}
\end{align}
So we can get
\begin{equation}
\Delta S_1=-4\pi m_1^2-8\pi m_1 (M-m_1).
\end{equation}
(Note that the minus sign has already been interpreted in sec. II.)
For the second emission of the particle with mass $m_2$,　the  entropy change of the black hole is
\begin{equation}
\Delta S_2=-4\pi m_2^2-8\pi m_2 (M-m_1-m_2).
\end{equation}
In a similar manner, we can get the change of the entropy that is taken away by the $i$-th particle,
\begin{equation}
\Delta S_i=-4\pi m_i^2-8\pi m_i (M- \overset{i}{\underset{j=1}{\sum}}m_j) .
\end{equation}
Therefore, after the black hole emitted $n$ particles the total  entropy change of the black hole is
\begin{align}
&\Delta S_{\text{total}}\equiv\overset{n}{\underset{i=1}{\sum}} \Delta S_{i}=    \nonumber \\
\begin{split}
=&-\overset{n}{\underset{i=1}{\sum}}4\pi m_i^2
-8\pi \overset{n}{\underset{i=1}{\sum}}m_i (M-\overset{n}{\underset{j=1}{\sum}}m_j)    \label{deltaS} -8\pi \overset{n}{\underset{i=1}{\sum}} \overset{n}{\underset{j=i+1}{\sum}}m_i m_j .
\end{split}
\end{align}
Here the first term in  Eq.(\ref{deltaS}) represents the self-entropy of all the emitted particles,
and the second term represents the correlations between the remaining black hole and the emitted particles,
and the third term represents the correlations between the emitted particles.
It is easy to obtain the relation
\begin{equation}
4\pi(M-\overset{n}{\underset{i=1}{\sum}}m_i)^2=4\pi M^2+\Delta S_{\text{total}},  \label{BH entropy:radiation}
\end{equation}
where $M-{\sum}_{i=1}^{n}m_i$ is the remnant mass of the black hole, and the whole left hand side  represents
the entropy of the  remaining black hole. Therefore,  Eq.(\ref{BH entropy:radiation})
shows that the entropy is conserved in Hawking radiation.
Repeating the above process of step by step analysis of each Hawking radiation until
the black hole is completely exhausted(${\sum}_{i=1}^{N}m_i=M$),
 we find that the entropy or information conservation is  preserved at all times,
\begin{align}
4\pi M^2&=-\Delta S_{\text{total}}\\
&=\overset{N}{\underset{i=1}{\sum}}4\pi m_i^2
+8\pi \overset{N}{\underset{i=1}{\sum}} \overset{N}{\underset{j=i+1}{\sum}}m_i m_j .
\label{BH entropy:radiation2}
\end{align}
Here the first term is the self-entropy of  all the radiation particles, and the second term
shows the correlations between the radiation particles.
Thus, we have proved that after a black hole is exhausted by Hawking radiation,
the entropy of the  black hole is carried away by the radiation particles and the correlations
between these particles.
On the other hand, because  we have shown that there are gravitational correlations between
these particles of Hawking radiation, these emissions must be correlated and
capable of carrying away information encoded within.

According to the tunneling method, the tunneling probability rate is
\begin{align*}
  \Gamma(M,m_1)\sim & \exp(\Delta S_1),\\
  =& \exp\left[-4\pi m_1^2-8\pi m_1(M-m_1)\right], \\
  =& \exp\left[-8\pi m_1(M-{m_1}/2)\right].
\end{align*}
This recovers the result of Parikh\&Wilczek\cite{pw00}. Obviously, it is not exactly thermal, which is different from the thermal case of a simple exponential $\Gamma \sim \exp(-8\pi M m_1)$.

Note that the current analysis is independent of the tunneling method, and hence the resolution of the paradox of black hole information loss given in \cite{zcyz09} indeed has a wider range of applicability. (C.f. also the analysis of information conservation in \cite{gc14} which is independent of  specific black-hole metrics.)

On the other hand, an important lesson we have learned from the tunneling method is that the back reaction is crucial for the conservation of energy. In fact, the back reaction not only makes radiation spectrum deviate from strict thermally, but changes the position of the horizon and hence the temperature of the black hole.
Now from the above analysis we see that the gravitational correlation entropy $S_c$ in \eqref{17} does not change if the radiated particles are translated by $\Delta x$, since it is independent of $\Delta x$. This is consistent with the results obtained in the case of temperature-varying holographic screens\cite{Wen14} where the mutual information generated by the translation of the radiated particle is restricted to zero for the entropy of holographic screens. In view of entropic gravity, this means that the effect induced by the translation of radiated particles are compensated by the change of temperature on the holographic screen, so that the entropic  force relation $\Delta S\propto\Delta r$ is preserved. Therefore, the above arguments for the entropy or information conservation hold at all times and no matter where the radiated particles have been translated.

\section{unifying the entropy of black hole and ordinary matter}
It is well know that the entropy is an extensive quantity for ordinary
matter\ in statistical mechanics, i.e. $S_{or}\propto N$ (or $S\propto M$,
where $N$ represent the particle number and $M$ is the mass of the ordinary
matter). However, the entropy of a black hole is no more an extensive
quantity, i.e. $S_{BH}\propto M^{2}$. This puzzle has troubled people for a long
time. In this section, we attempt to solve this problem by comparing their gravitational correlation entropies, and demonstrate how ordinary matters
become a black hole.

Assume that a spherical system with mass $M$ consisting of $N$ identical
particles each of which has mass $m$, and the average distance
between the particles is $l$. We further assume that all the particles are uniformly
distributed in the system for simplicity. Then the radius of the system is
$nl$, the volume $V$ and density of mass $\rho$ are obtained as
\begin{align*}
N  &  =4\pi n^{3}/3,\quad V=Nl^{3}=4\pi n^{3}l^{3}/3,\\
M  &  =Nm,\quad\rho=\frac{M}{V}=\frac{m}{l^{3}}.
\end{align*}
The entropy of the system should include the entropy of every particle $S_{m}$
and the correlation entropy $S_{c}$ which is due to the gravitational interactions
between these particles. Suppose that these particles are located on a lattice each site of which has 3-dimensional spatial coordinates $(i,j,k)$. Then the total entropy of the system is
\begin{equation}
S_{N}=NS_{m}+\overset{n}{\underset{i,j,k}{\sum}}S_{(i,j,k)c}(N)  \label{sn}%
\end{equation}
where $S_{(i,j,k)c}(N)$ represents the correlation entropy of the $(i,j,k)$-th
particle that is associated with the other $N-1$ particles. It is obvious that
$S_{(i,j,k)c}(N)\leq S_{(0,0,0)c}(N)\equiv S_{oc}(N)$, ($i^{2}+j^{2}+k^{2}%
\neq0$) for spherical objects. So%
\begin{equation}
\overset{n}{\underset{i,j,k}{\sum}}S_{(i,j,k)c}(N)\leq NS_{oc}(N)\label{s<so}%
\end{equation}
Because it is difficult to calculate $S_{(i,j,k)c}(N)$ for every particle, we
will first calculate this bound $S_{oc}(N)$,
\begin{align*}
S_{oc}(N) &  =-\underset{i,j,k}{\sum}\int_{\infty}^{l_{ijk}}8\pi m^{3}%
\frac{dx}{x^{2}}=8\pi m^{3}\underset{i,j,k}{\overset{N}{\sum}}\frac{1}{l_{ijk}}\equiv8\pi m^{3}\frac{C(n)}{l}.
\end{align*}
To calculate $C(n)$, we suppose that  the distance between the particle $(i,j,k)$ and the coordinate origin of the lattice
is $l_{ijk}=l\sqrt{i^{2}+j^{2}+k^{2}}$(where $i,j,k$ are integers). Then we
have%
\begin{equation}
C(n)=\underset{i^{2}+j^{2}+k^{2}\leq n^{2}}{\sum}\frac{1}{\sqrt{i^{2}%
+j^{2}+k^{2}}},\quad(i^{2}+j^{2}+k^{2}\neq0)\label{cn}.
\end{equation}
When $n\gg1$,
\[
C(n)-C(n-1)  =\left[  N(n)-N(n-1)\right]  /n\approx4\pi n.
\]
That is,
\begin{align}
\frac{dC(n)}{dn} &  \approx4\pi n,\nonumber\\
\Rightarrow\quad C(n) &  \approx2\pi n^{2}\label{cn1}.%
\end{align}
Then we have,
\begin{equation}
S_{oc}(N)  =4m^{3}(6\pi N)^{2/3}/l =16\pi^{2}m^{3}n^{2}/l\label{scn2}.%
\end{equation}

Now, we will compare the correlation entropy $S_{c}(N)$ with $S_{m}$.
Suppose that the particle $m$ constituted by two smaller particles with equal mass
$m/2$ and the distance between them is \thinspace$l_{m}$, then $S_{m}$  consists of two parts
as before. The  first part is the self-entropy of the two subparticles $2S_{m/2}$
and the second part is the correlation entropy between the two subparticles
$2\overset{\sim}{S_{c}}$. Therefore, we can get%
\begin{align}
S_{m} &  =2S_{m/2}+2\overset{\sim}{S_{c}}\nonumber,\\
&  =2S_{m/2}+2\pi m^{3}/l_{m}\label{sm2}.%
\end{align}
Using Eq. (\ref{s<so})  and Eq.(\ref{sm2}), we have%
\begin{equation}
\frac{S_{c}(N)}{S_{m}}<\frac{16\pi^{2}m^{3}n^{2}/l}{2S_{m/2}+2\pi m^{3}/l_{m}
}.\label{sc/sm}%
\end{equation}
Up to now, we still do not know $S_{m/2}$. But obviously, it should be positive, so we
have,
\[
\frac{S_{c}(N)}{S_{m}}<\frac{8\pi n^{2}l_{m}}{l}.%
\]
For ordinary matter, $l_{m}/l\ll1$, therefore, as long as the volume (related
to $n$) is not very large, we have
\begin{equation}
\frac{S_{c}(N)}{S_{m}}<\frac{8\pi n^{2}l_{m}}{l}\ll1.\label{sc/sm1}%
\end{equation}
Therefore, for ordinary matter the gravitational correlation entropy between the particles, the second term in  Eq.(\ref{sn}), can
be ignored. Then Eq.(\ref{sn}) becomes
\begin{equation}
S_{N}\approx NS_{m}.\label{sn2}%
\end{equation}
This shows that the entropy of ordinary matter is proportional to the number
of particles. It is an extensive quantity because of the gravitational  correlation entropy of ordinary matter caused by gravitational interactions between these  particles is too small and can be ingored.

On the other hand, note that Eq.(\ref{scn2}) shows that the correlation entropy is not convergent as
 particle number of the system increases. That is to say, even
though the density of the system is very small, $S_{c}(N)$ can be bigger than
$S_{m}$ as long as the particle\ number of the system is  large enough. At this
point, the entropy of this system $S_{N}$ is no more an extensive quantity. The black hole entropy is an example of this case, since the large mass of a black hole ensures the large number of particles and a very small $l$.

Therefore, we have constructed a possible unified description of both extensive and nonextensive entropies. The characterization is given by the gravitational correlation entropy of the interior matters for different particle numbers  at different length scales.

\section{Conclusion and discussions}

In this paper we introduced the concept of gravitational correlation entropy, which measures the correlations between gravitationally interacting objects. With this concept, we first shown that the entropy of a Schwarzchild black hole originates from the gravitational correlations between the interior matters of the black hole. Then we calculated the correlation entropy between Hawking radiation particles to show the information conservation of Hawking radiation in a wider context than in \cite{zcyz09}. Finally, we gave an attempt to unify the nonextensive black hole entropy with the extensive thermal entropy by comparing the magnitude of the correlation entropy within them.

Some final remarks are in order:
\begin{enumerate}
\item The entropic gravity, in spite of its controversial interpretations (see, for example, \cite{Gao11}), should play a  fundamental role in clarifying various confusions in gravitation. This can be corroborated by the recent progress that solutions to gravitational field equations can be reconstructed solely from the spacetime thermodynamics\cite{ZHZL14} where the key concept of the reconstruction, the mass form of Misner-Sharp on a hypersurface, plays a similar role of the holographic screen as in the entropic gravity.

\item With the help of the gravitational correlation entropy, we have successfully  recovered the information conservation arguments given in \cite{zcyz09}. The analysis of \cite{zcyz09} is performed in the tunneling formalism of Hawking radiation, while in this paper the analysis works in a wider context. (C.f. also the metric-independent arguments in \cite{gc14}). Besides the tunneling formalism, a common attack on the resolution of \cite{zcyz09} is the arbitrariness in partition the quanta of Hawking radiation, since the chain rule for conditional entropy is always satisfied for any partition regardless of the dynamics. In our discussion in sec. II, we see that such an arbitrariness is a blessing rather than a problem. After a black hole is disintegrated into Hawking radiations and a possible remnant, as long as the mutual information of this general process is positive, then in view of the noiseless channel coding theorem\footnote{This theorem also can used to explain the laws of black hole thermodynamics. See, e.g. \cite{guo15}.}, it requires additional resources outside the black hole to store these additional information. In other words, the information has indeed come out from the black hole and the correlation is not necessarily those between radiated particles. This is why such an argument for information conservation still hold for extremal black holes\cite{CSY16}. This proposal assures us that the analysis of \cite{zcyz09} is simple but information-theoretically profound.

\item The translation invariance of the correlation entropy ensures the unitarity of translation operators. In \cite{AN13} the gravitational correlation entropy, or the mutual information in the tunneling formalism, is included into the non-extensive Tsallis entropy formula. This poses a problem that which form of entropy is more fundamental. Since the entropic force of E. Verlinde\cite{ev11} is a kind of reverse of the line of thoughts of the traditional gravitational researches, it is hard to tell which one is more fundamental. A physically sounder approach is to test such entropies in different situations by different formalisms. In this paper we have attempted some tests. Another promising way of describing the translations (though quite different) and holographic screens is the more delicate notions of BMS super translations on the black hole horizon, the relevance of which to the black hole information problem is recently investigated in \cite{HPS16}.

\item Our analysis is restricted to the simplest case of Schwarzschild black hole, since it has the clearest entropic force explanations. In order to consider other types of black holes, one can either (i) interpret degrees of freedom other than gravity (see, for examples, coulomb force\cite{Wang10} and rotation\cite{Cu11}) as entropic forces, or (ii) do direct calculations in the old framework to get a modified picture with corrections to relevant quantities\cite{LWW10}. In the latter approach, there will be a failure of the entropic idea of gravity if the noncommutative effects are present\cite{MK12}. This is not surprising, because in the analysis of information conservation of noncommutative black holes, the noncommutativity of spacetime contributes an additional term to the mutual information (or gravitational correlation entropy) other than the term from the commutative case\cite{zczy11}. This suggests that the noncommutative effects of (quantum) spacetime are quite different from the corrections obtained from entropic gravity. (C.f. \cite{GP13} for a recent discussion on this point.)
\end{enumerate}
\begin{acknowledgments}
We are deeply indebted to Xiao-Kan Guo for numerous suggestions and criticisms on earlier versions of this paper. This work is supported in part by NSFC No. 61471356.
\end{acknowledgments}


\begin{thebibliography}{9}
\bibitem {ev11} E. Verlinde, J.High Energy Phys. {\bf1104}, 029 (2011).

\bibitem{swh71} S. W. Hawking, Phys. Rev. Lett. {\bf26}, 1344 (1971).

\bibitem{jdb73} J. D. Bekenstein, Phys. Rev. D {\bf7}, 2333 (1973).

\bibitem{swh74} S. W. Hawking, Nature {\bf284}, 30 (1974).

\bibitem{swh75} S. W. Hawking, Commun. Math. Phys. {\bf43}, 199 (1975).

\bibitem{rl91} R. Landauer, Phys. Today {\bf44}, 23 (1991).

\bibitem{nc00} M. A. Nielsen, I. L. Chuang, {\it Quantum Computation and Quantum Information}, (Cambridge University Press, Combridge, 2000).
\bibitem{swh76} S. W. Hawking, Phys. Rev. D {\bf14}, 2460 (1976).

\bibitem{BC11} V. Balasubramanian, B. Czech, Class. Quantum Grav. {\bf28}, 163001 (2011).

\bibitem{Har14} D. Harlow, arXiv:1409.1231.

\bibitem{zcyz09} B.-c. Zhang, Q.-y. Cai, L. You, M.-S. Zhan, Phys. Lett. B {\bf675}, 98 (2009).

\bibitem{BKLS86} L. Bombelli, R. K. Koul, J. Lee, R. D. Sorkin, Phys. Rev. D {\bf34}, 373 (1986).

\bibitem{ms93} M. Srednicki, Phys. Rev. Lett. {\bf71}, 666 (1993).

\bibitem{fn93} V. P. Frolov, I. Novikov, Phys. Rev. D {\bf48}, 4545 (1993).

\bibitem{Lee} T.-D. Lee, {\it Statistical Mechanics}, (Shanghai Publishing House for Science and Technology, Shanghai, 2006).

\bibitem{gsl08} S.-J. Gu, C.-P. Sun, H.-Q. Lin, J. Phys. A: Math. Theor. {\bf41} 025002 (2008).

\bibitem{Gao11} S. Gao, Entropy, {\bf13}, 936 (2011).

\bibitem{wu76} W. Unruh, Phys. Rev. D,  {\bf14}, 870 (1976).

\bibitem{Kob11} A. Kobakhidze, Phys. Rev. D {\bf83}, 021502 (2011).

\bibitem{AN13} E. M. Abreu, J. A. Neto, Phys. Lett. B {\bf727}, 524 (2013).

\bibitem{Zhang15} B.-c. Zhang, Phys. Rev. D {\bf92}, 081501(R) (2015).

\bibitem{Sah11} H. Sahlmann, Phys. Rev. D {\bf84}, 104010 (2011).

\bibitem{pw00} M. K. Parikh, F. Wilczek, Phys. Rev. Lett. {\bf85},5042(2000).

\bibitem{gc14} X.-K. Guo, Q.-y. Cai, Int. J. Theor. Phys. {\bf53}, 2980 (2014).

\bibitem{Wen14} W.-Y. Wen, Eur. Phys. J. C {\bf74}, 2912 (2014).

\bibitem{ZHZL14} H.-s. Zhang, S. A. Hayward, X.-H. Zhai, X.-Z. Li, Phys. Rev. D {\bf89}, 064052 (2014).
\bibitem{guo15} X.-K. Guo, arXiv:1512.05277.

\bibitem{CSY16} Q.-y. Cai, C.-P. Sun, L. You, Nucl. Phys. B {\bf905}, 327 (2016).

\bibitem{HPS16} S. W. Hawking, M. J. Perry, A. Strominger, Phys. Rev. Lett. {\bf116}, 231301 (2016).

\bibitem{Wang10} T. Wang, Phys. Rev. D {\bf81}, 104045 (2010).

\bibitem{Cu11} A. Curir, Commun. Theor. Phys. {\bf55}, 594 (2011).

\bibitem{LWW10} Y.-X. Liu, Y.-Q. Wang, S.-W. Wei, Class. Quantum Grav. {\bf27}, 185002 (2010).

\bibitem{MK12} S. H. Mehdipour, A. Keshavarz, EPL, {\bf98}, 10002 (2012).

\bibitem{zczy11} B.-c. Zhang, Q.-y. Cai, M.-S. Zhan, L. You, EPL, {\bf94}, 20002 (2011).

\bibitem{GP13} C. M. Gregory, A. Pinzul, Phys. Rev. D {\bf88}, 064030 (2013).


\end{thebibliography}
\end{document}